\documentclass[aps,prc,preprintnumbers,fleqn,
superscriptaddress,floatfix]{revtex4-1}
\usepackage{amsmath,amsfonts,bm}
\usepackage[dvipdfm]{graphicx}
\usepackage[dvips]{color}

\begin{document}

\preprint{INHA-NTG-08/2010}

\title{Binding energy per nucleon and hadron properties in nuclear
 matter}

\author{Ulugbek Yakhshiev }
\email{yakhshiev@inha.ac.kr}

\affiliation{Department of Physics, Inha University, Incheon
402-751, Republic of Korea }

\author{Hyun-Chul Kim}
\email{hckim@inha.ac.kr}

\affiliation{Department of Physics, Inha University, Incheon
402-751, Republic of Korea }

\date{February, 2011}

\begin{abstract}
We investigate the binding energy per nucleon and hadron properties in
infinite and homogeneous nuclear matter within the framework of the
in-medium modified Skyrme model. We first consider the medium
modifications of the single hadron properties by introducing the optical
potential for pion fields into the original Lagrangian of the Skyrme
model. The parameters of the optical potential are well fitted to
the low-energy phenomenology of pion-nucleus scattering. Furthermore,
the Skyrme term is also modified in such a way that the model
reproduces bulk properties of nuclear matter, in particular, the
binding energy per nucleon. The present approach is self-consistent:
the single hadron properties in a nuclear medium, their effective
in-medium interactions, and the bulk matter properties are treated
on the same footing.
\end{abstract}

\pacs{12.39.Dc, 21.65.Ef, 21.65.Jk}

\keywords{Skyrme model, binding energy, symmetry energy, $\pi NN$
  coupling constant}

\maketitle

\section{Introduction}
The equation of state (EOS), which gives the density
dependence of the binding energy per nucleon for a given nucleus, has
been one of the most importance issues in nuclear many-body problems.
There is a great amount of different theoretical approaches in trying
to describe the EOS and, clearly, we can mention only some of few
representatives~\cite{Chabanat:1997qh,Chabanat:1997un,TerHaar:1986ii, 
Brockmann:1990cn,Hofmann:2000vz,Serot:1997xg,Serot:1984ey}. In
general, those approaches and corresponding representatives  can be
classified into three classes as microscopic many-body 
approaches~\cite{Chabanat:1997qh,Chabanat:1997un,TerHaar:1986ii,
Brockmann:1990cn}, effective field
theories~\cite{Hofmann:2000vz,Serot:1997xg}, and
phenomenological methods~\cite{Serot:1984ey}. These works for
many-body problems provide very useful tools for understanding
properties of dense and hot matter.

On the other hand, the Skyrme model~\cite{Skyrme:1962vh,Adkins:1983ya}
also presents a simple but good framework for investigating bulk
properties of nuclear matter.  In general, one can classify various
Skyrmion approaches into two subclasses: In the first one
investigations are mainly devoted to the classical crystalline
structure and its behavior under the extreme
conditions~\cite{Klebanov:1985qi,Lee:2003aq}. In the second approach
the properties of exotic many-baryon systems were
treated~\cite{Battye:2006na,Manton:2006tq,Manko:2006dr}. There are
also some early attempts to explain many-body
systems within the Skyrme model~\cite{Manton:1986pz,Jackson:1988ji},
considering the single skyrmion in hypersphere.

Moreover, there is an another alternative way to examine properties of nuclear
matter within the Skyrme model, i.e. to study the properties of the
single skyrmion in nuclear matter~\cite{Rakhimov:1996vq}. Furthermore,
the medium-modified Skyrme
model~\cite{Rakhimov:1996vq,Rakhimov:1998hu}, in connection with
quantum-mechanical variational methods~\cite{Diakonov:1987zp} (like
the Hartree-Fock method), can be applied to the analysis of the bulk
properties of nuclear matter~\cite{Yakhshiev:2004pj}. To perform this
analysis, it is essential to know the properties of the single hadron
and the $NN$ interaction in a
symmetric~\cite{Rakhimov:1996vq,Rakhimov:1998hu} as well as
asymmetric~\cite{Meissner:2007ya}) nuclear environment. Consequently,
the behavior of the hadrons in nuclear medium must be taken into
account. However, the variational calculations can only estimate the
upper boundary value of minimized quantities. Thus, one may still
consider how to improve the results.

As a more realistic approach, the in-medium modified Skyrme
model~\cite{Rakhimov:1996vq} itself can be used so as to reproduce the properties of the
single hadron in nuclear matter as well as those of matter in bulk.
This is the aim of the present work. To carry out the proposed goal,
we consider not only the changes of the kinetic and mass terms of the
standard Skyrme Lagrangian as done in Ref.~\cite{Rakhimov:1996vq} but
also possible modifications of the Skyrme term. It is well known that
Skyrme's quartic stabilizing term may be related to vector
mesons~\cite{Meissner:1987ge} that can be realized in implicit gauge
symmetry of the nonlinear sigma model
Lagrangian~\cite{Bando:1984ej}. In this sense, the modification of the
Skyrme parameter may be pertinent to the changes of the vector mesons
in nuclear matter.

The present work is organized as follows: in the next Section, we
explain how the original Skyrme model can be modified in medium.
In Section III, we present the corresponding numerical results and
discuss them. In particular, we show that the quartic Skyrme term can
be modified in such a way that it minimizes the binding energy per
nucleon. We discuss the changes of the mass splitting of the nucleon
and the $\Delta$ isobar and those of the $\pi NN$ coupling constant.
We also estimate the symmetry energy in the Weizs\"acker-Bethe-Bacher
formula. The last Section is devoted to summary and outlook of this
work.

\section{Medium modification of hadron properties
\label{sec:Lag}}
In Ref.~\cite{Rakhimov:1996vq}, the in-medium modified Skyrme
Lagrangian was presented, the mass term being modified based on the
phenomenology of low-energy pion-nucleus scattering. The modified mass
term leads also to the changes of the kinetic term. In the present
work, we additionally consider the modification of the Skyrme
stabilizing term.  The resulting Lagrangian is given as follows:
\begin{eqnarray}
{\cal L}^*&=&\frac{F_\pi^2}{16}\,{\rm Tr}\left(\frac{\partial
U}{\partial t}\right)\left(\frac{\partial U^\dagger}{\partial
t}\right) 
-\frac{F_\pi^2}{16}\,\alpha_p({\bm r}){\rm
  Tr}({\bm\nabla} U)\cdot({\bm\nabla}
 U^\dagger)
+\frac{1}{32e^2\gamma({\bm r})}\,{\rm
Tr}[U^\dagger\partial_\mu U,U^\dagger\partial_\nu U]^2\cr
&& +\frac{F_\pi^2m_\pi^2}{16}\,\alpha_s({\bm r}){\rm Tr}(U+
U^\dagger-2)\,, \label{Eq:Lag}
\end{eqnarray}
where $F_\pi$ denotes the pion decay constant, $e$ is the Skyrme
parameter, and $m_\pi$ stands for the pion mass. The medium
functionals $\alpha_s$ and $\alpha_p$ are written in the following
forms
\begin{eqnarray}
\alpha_s=1-\frac{4\pi b_0\rho({\bm r})f}{m_\pi^{2}},\quad
\alpha_p=1-\frac{4\pi c_0\rho({\bm r})}{f+g_0'4\pi
c_0\rho({\bm r})},\label{medfunc}
\end{eqnarray}
which represent the influence of the surrounding environment on the
properties of the single skyrmion. These parameters are related
respectively to the corresponding phenomenological $S$- and $P$-wave
pion-nucleus scattering lengths and volumes, i.e. $b_0$ and $c_0$,
and describe the pion physics in a baryon-rich
environment~\cite{Ericsonbook}. The density of the surrounding
nuclear environment is given by $\rho$, $g_0^{\prime}$ denotes the
Lorentz-Lorenz or correlation parameter, $f=1+m_\pi/m_{N}^{\rm free}$
represents the kinematical factor, and $m_{N}^{\rm free}$ is the
nucleon mass in free space. In addition to these changes of the mass
and kinetic terms done in  Ref.~\cite{Rakhimov:1996vq}, we introduce
the new density-dependent functional $\gamma({\bm r})=\gamma(\rho({\bm
  r}))$ which provides the in-medium dependence of the Skyrme
parameter, i.e. $e^2\to e^{*2}=e^2\gamma$. As mentioned in
Introduction, this medium modification can be related to the density
dependence of vector meson properties. Furthermore, to fix this
additional dependence in the present work, we will concentrate on
the bulk properties of infinite and homogenous nuclear matter with
constant density ($\rho=\mbox{const}$). Note that the Lagrangian
(\ref{Eq:Lag}) is modified in such a way~\cite{Rakhimov:1996vq} that
at zero density it reduces to the original Lagrangian of the Skyrme
model and at the linear approximation it reproduces the well-known
equation of the pion fields in nuclear medium~\cite{Ericsonbook}.

In order to analyze the field equation for the classical pion field in
homogenous nuclear matter one can choose the spherically symmetric
``hedgehog" form for the boson field
$U=\exp\{i\hat{{\bm n}}\cdot{\bm \tau}F(r)\}$, where
${\bm n}$ denotes the unit vector in coordinate space,
${\bm \tau}$ are the usual Pauli matrices, and $F(r)$ stands for the
profile function of the pion field. The pertinent field equation,
which is given as
\begin{equation}
F^{\prime\prime}(x) \left(\alpha_p\gamma x^2+8s^2\right)+2\alpha_p\gamma
xF'(x)-\alpha_s\gamma \beta^2x^2s
+\left(4F'(x)^2-\alpha_p\gamma-\frac{4s^2}{x^2}\right)
\sin(2F)\;=\;0,\label{feq}
\end{equation}
is obtained by minimizing the medium-modified mass of the static
skyrmion
\begin{equation}
M^*_S \;=\; \frac{\pi
F_\pi}{e}\int_0^\infty  dx
\left\{\alpha_p\left(\frac{x^2F^{\prime2}}{2} +
  \sin^2 F\right) +\frac{4\sin^2
  F}{\gamma}\left(F'^2+\frac{\sin^2F}{2x^2}\right)
+\alpha_s\beta^2x^2(1-\cos F)\right\},\quad
\end{equation}
In the last two expressions, we have introduced the
dimensionless variable $x=eF_\pi r$ and the new constant
$\beta=m_\pi/(eF_\pi)$.

The collective quantization of the classical
skyrmion~\cite{Adkins:1983ya} yields the in-medium modified nucleon
mass and the  corresponding $\Delta- N$ mass splitting respectively
as
\begin{eqnarray}
m_{ N}^*&=& M^*_S+\frac{3}{8\lambda^*},\qquad m_{\Delta-
N}^*\,=\,\frac{3}{2\lambda^*},\nonumber\\
\lambda^*&=&\frac{2\pi}{3e^3F_\pi} \int_0^\infty dx \,x^2\sin^2
F \left\{1+
\frac{4}{\gamma}\left(F'^2+\frac{\sin^2 F}{x^2}\right)\right\} \,,
\label{inmom}
\end{eqnarray}
where $\lambda^*$ denotes the in-medium moment of inertia of the
skyrmion. The meson-baryon vertices in nuclear matter can be derived
by calculating the in-medium modified $\pi NN$ form
factor~\cite{Rakhimov:1998hu}
\begin{equation}
G_{\pi NN}^*(q^2) \;=\; \frac{4\pi  M_{ N}^* }{3e^2F_\pi}\,
\alpha_p\int_0^\infty\frac{j_1({\tilde q}x)}{{\tilde q}x}\,S_\pi(x)x^3
{\rm d}x\,,
\end{equation}
where ${\tilde q=q/eF_\pi}$, $j_1(qx)$ is the spherical Bessel
function with order 1 and $S_\pi(x)$ is defined as
\begin{equation}
S_\pi(x) \;=\; -\left({2x^{-1}F'}+F^{\prime\prime}\right)\cos F
+\left(F'^2+{2}{x^{-2}}+{\alpha_s}{\alpha_p^{-1}}\,m_\pi^2\right)\sin
F \,.
\end{equation}

Using the Lagrangian given in Eq.~(\ref{Eq:Lag}), one can calculate
the in-medium modifications of the single nucleon properties and the
pion-nucleon coupling constant. Note that the parameters of the model
are fitted to be $F_\pi=108.78$~MeV and $e=4.85$ so as to reproduce
the experimental values of the nucleon and $\Delta$ in free space.
Consequently, the pion mass is also fixed to be its experimental value
for the neutral pion, i.e. $m_\pi=134.98$~MeV. A set of values of
parameters in the medium functionals~(\ref{medfunc}) are taken from
the analysis of phenomenological data for pion-nucleus
scattering~\cite{Ericsonbook}.

\section{Numerical results and Discussions}
As is clear from the discussions in the previous Section,
if the modification of the Skyrme term is ignored, the values of all
input parameters are fitted to the phenomenology or taken from
it. When, however, the Skyrme term is modified, we introduce one
additional functional $\gamma(\rho)$ which, in general, may be related
to the vector meson properties in nuclear matter. One can also note
that the lessening value of the Skyrme parameter in nuclear
medium may correspond to a decrease of the $g_{\rho \pi\pi}$ coupling
and, therefore, to the change of the rho meson width in nuclear matter
or to a diminishing value of its mass in medium,
i.e. $m_\rho^*/m_\rho<1$. There are experimental indications to those
changes of the $\rho$ meson
properties~\cite{Ozawa:2000iw,Naruki:2005kd,Weygand:2007ky} and the
theoretical predictions~\cite{Brown:1991kk,Hatsuda:1991ez}. Following
the ideas presented in those theoretical approaches, one may be able
to fit $\gamma$. However, it is still under debate how the properties
of the $\rho$ meson undergo in medium both theoretically and 
experimentally, and model-dependent. Thus, in the present
work, we will proceed to fit the 
form of the functional $\gamma$ to the bulk properties of nuclear
matter rather than following a specific model.

As a first step, we explicitly choose its form to
reproduce the first coefficient (volume term) in the semiempirical
Weizs\"acker-Bethe-Bacher mass formula. Then the binding energy per
nucleon at a given density can be defined simply as
\begin{equation}
\Delta E_{B=1}=m_{ N}^*(\rho)-m_{ N}^{\rm free}\,.
\label{BE}
\end{equation}
This is somehow a crude approximation but a comprehensive one.
One could even fit the form of $\gamma$ by investigating different
terms in the mass formula and by examining the interplay between
them. However, within the present work, the approximation defined in
Eq.~(\ref{BE}) will be enough for the qualitative analysis of the
changes due to the modification of the Skyrme parameter.

In the present calculation we have tried various forms of the
dependence of $\gamma$ on the nuclear density $\rho$ such as linear,
quadratic, polynomial, exponential forms, etc. It turns out that the
best fit to the ground state of nuclear matter is achieved by
the following form
\begin{equation}
\gamma(\rho)=\exp\left(-\frac{\gamma_{\rm num}\rho}{1+\gamma_{\rm
den}\rho}\right)\,,
\label{gamma}
\end{equation}
where $\gamma_{\rm num}$ and $\gamma_{\rm den}$ are variational
parameters.
\begin{figure}[htb]
\centerline{\includegraphics[scale=0.8]{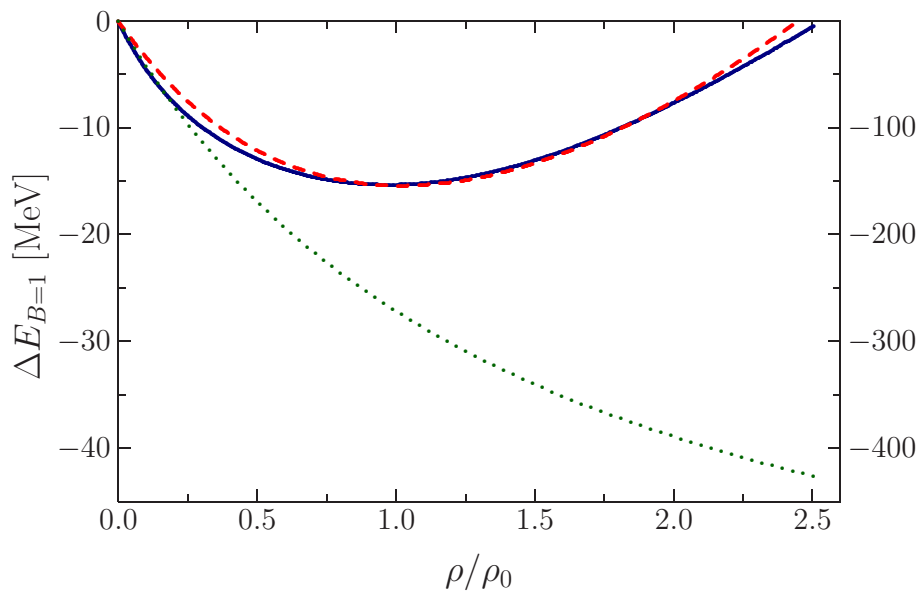}}
\caption{(Color online). The binding energy per nucleon as a function of
$\rho/\rho_0$. The solid curve (left scale) corresponds to the
parametrization of $\gamma$ in Eq.~(\ref{gamma}), with $\gamma_{\rm
num}=2.1m_\pi^{-3}$, $\gamma_{\rm den}=1.45m_\pi^{-3}$ and $P$-wave
scattering volume $c_0=0.21m_\pi^{-3}$ used.
The dashed one (left scale) draws the case when $\gamma_{\rm
num}=0.8m_\pi^{-3}$, $\gamma_{\rm den}=0.5m_\pi^{-3}$ and $P$-wave
scattering volume $c_0=0.09m_\pi^{-3}$. $S$-wave scattering length is
fixed at $b_0=-0.024m_\pi^{-1}$ and the correlation parameter has
value $g'=0.7$. The dotted one (right scale) shows the case that
Skyrme term intact in nuclear matter and consequently
$\gamma(\rho)=1$. The normal nuclear density is given as
$\rho_0=0.5m_\pi^3$.}
 \label{fig:1}
\end{figure}
Let us first discuss the behavior of the binding energy when the
Skyrme term is intact in nuclear matter, i.e. $\gamma(\rho)=1$. The
corresponding binding energy is depicted as the dotted curve in
Fig.~\ref{fig:1} with the energy scale drawn at the
right vertical axis. One can note that in this case the binding energy
monotonically falls off as the density increases. In the language of
the single skyrmion, it indicates that the skyrmion swells to a larger
volume and all skyrmions of the system start to overlap. Thus, the
density of the system continuously increases. This is not surprising,
because the medium modification in this case can be simply related to
that of the pion decay constant $F_\pi\rightarrow
F_\pi^*=F_\pi\sqrt{\alpha_p}$. For the moment, one can ignore the
explicit chiral symmetry breaking term in the Lagrangian, because its
influence to the stability is rather small in comparison with the
effects coming from the first two terms. The decreasing value of the
pion decay constant changes the contribution from the
nonlinear kinetic term. As a result the skyrmions swell to the larger
volume and it is necessary to prevent this by some mechanism. It
implies that one must introduce either strong repulsive $NN$
interactions at short distances or some mean-field mechanism as in the
Walecka model~\cite{Serot:1997xg,Serot:1984ey}. However, one
interesting way to avoid this collapse may be to modify the Skyrme
term and it is also physically motivated as discussed at the beginning
of this Section. Moreover, it is much simpler and transparent to
consider the modification of the Skyrme term.

We also present two different results with the modified
Skyrme term in Fig.~\ref{fig:1}: The solid and dashed curves
draw the parametrization of Eq.~(\ref{gamma}) with the energy scale
depicted at the left vertical axis. Note that the values of the
variational parameters, $\gamma_{\rm num}$ and $\gamma_{\rm den}$, are
chosen in a such way that the minimum of the binding energy occurs at
the normal nuclear matter density and reproduces correctly the first
coefficient in the empirical formula for the binding energy. The
difference between these two curves is due to the fact that we use two
different values of the $P$-wave scattering length, i.e.
$c_0=0.21m_\pi^{-3}$ for the solid curve and $c_0=0.09m_\pi^{-3}$ for
the dashed one. The results show that the dependence on the density is
rather insensitive to the changes of input parameters from
pion-nucleus scattering phenomenology and moreover the effect of
the changes in $b_0$ is even milder.

In order to see the validity of Eq.~(\ref{gamma}), it is of great
importance to examine the changes of other physical observables.
Let us first discuss thermodynamic properties of nuclear
matter. The pressure is given by the following formula
\begin{equation}
p=\rho\,\frac{\partial \epsilon}{\partial\rho}-\epsilon=
\rho^2\frac{\partial \Delta E_{B=1}}{\partial \rho}\,,
\end{equation}
where $\epsilon$ is the total binding energy of nuclear matter
per unit volume. It vanished naturally at the equilibrium point and
for the parametrization of Eq.~(\ref{gamma}). We want to emphasize
that the pressure is always decreasing with the Skyrme term intacted.

Another important quantity is the compressibility of nuclear matter
expressed as
\begin{eqnarray}
K&=&9\rho_0^2\left.\frac{\partial^2\Delta E_{B=1}}{\partial
    \rho^2}\right|_{\rho=\rho_0} 
=9\rho_0^2\left\{\frac{\partial^2\alpha_s}{\partial
    \rho^2}\right|_{\rho=\rho_0} \frac{\pi
F_\pi}{e}\int_0^\infty dx \, \left(\frac{x^2F'^2}2 +
  \sin^2 F\right) \\
&+&\frac{\partial^2}{\partial \rho^2}
\left.\left(\frac1\gamma\right)\right|_{\rho=\rho_0} \left[\frac{4\pi
F_\pi}{e}\int_0^\infty dx\,
\left(F'^2+\frac{\sin^2F}{2x^2}\right)\sin^2 F
\right. 
- \left.\left.
\frac{\pi}{e^3F_\pi\lambda^{*2}} \int_0^\infty dx\, x^2
\left(F'^2+\frac{\sin^2 F}{x^2}\right)\sin^2 F\right]\right\},\nonumber
\end{eqnarray}
and the corresponding results are listed in Table~\ref{tab1} with two
different values of the $S$-wave scattering length
$b_0$~\cite{Ericsonbook} used.
\begin{table}[h]
\begin{center}
\begin{tabular}{c|c|c|c|c|c}\hline\hline
$b_0\,[m_\pi^{-1}$] & $c_0\,[m_\pi^{-3}]$&$\gamma_{\rm
  num}\,[m_\pi^{-3}]$ & $\gamma_{\rm den}\,[m_\pi^{-3}] $& $K$ [MeV]
& $m_{N-\Delta}^*$ [MeV] \\ \hline
$-0.024$ & $0.21$& $2.098$&$1.451$&$1647.47$&$105.21$\\
$-0.024$ & $0.15$& $1.448$&$0.998$&$1148.18$&$129.39$\\
$-0.024$ & $0.09$& $0.797$&$0.496$&$~~582.79$&$170.34$\\ \hline
$-0.029$ & $0.21$& $2.106$&$1.506$&$1637.16$&$107.13$\\
$-0.029$ & $0.15$& $1.444$&$1.031$&$1142.00$&$131.59$\\
$-0.029$ & $0.09$& $0.785$&$0.502$&$~~580.03$&$172.91$\\
\hline\hline
\end{tabular}
\end{center}
\caption{Compressibility of nuclear matter $K$ and an effective
  $\Delta$-nucleon mass difference $m_{N-\Delta}^*$ at the normal
  nuclear matter density $\rho_0=0.5m_\pi^3$. The variational
  parameters $\gamma_{\rm num}$ and $\gamma_{\rm den}$ are fitted to
  reproduce the minimum of the binding energy per nucleon $\sim
  15.7$~MeV at the normal nuclear matter density. The correlation
  parameter is taken to be $g'=0.7$.\label{tab1}}
\end{table}
The results show that the compressibility of nuclear matter and the
effective $\Delta-N$ mass difference are rather stable under the
change of $b_0$. On the contrary, they are quite sensitive to the
value of $P$-wave scattering volume $c_0$. At the empirical value of
$c_0=0.21m_\pi^{-3}$, the compressibility turns out to be very large
($K\sim 1640$~MeV) in comparison with those obtained in relativistic
Dirac-Brueckner-Hartree-Fock
approaches~\cite{TerHaar:1986ii,Brockmann:1990cn} and in the Walecka
model~\cite{Serot:1984ey}. We find that as lower values of $c_0$ are
used $K$ is noticeably decreased. For example, for $c_0=0.09m_\pi^{-3}$
the compressibility is already consistent with that of the Walecka
model ($K\sim 580$~MeV). If one uses even a smaller value of $c_0$
such as $c_0=0.06m_\pi^{-3}$, the result of $K$ is further brought down
to be comparable with that in Dirac-Brueckner-Hartree-Fock approaches
($K\sim 300$~MeV), which is close to the empirical value. It indicates
that the present work prefers smaller values of $c_0$ than that used
in the pionic atom analysis as far as the compressibility is
concerned. Note that a similar conclusion about $c_0$ was drawn from
the analysis of an effective axial-vector coupling within the original
medium-modified Skyrme model~\cite{Rakhimov:1996vq}. We remind that
$K$ is sensitive to the position of the saturation point. Fitting the
saturation point at slightly lower densities, we see that the
compressibility decreases drastically. However, the situation may
change if one considers a more accurate approximation with the surface
and symmetry energy terms explicitly taken into account in
Eq.~(\ref{BE}). 

\begin{figure}[htb]
\centerline{\includegraphics[scale=0.8]{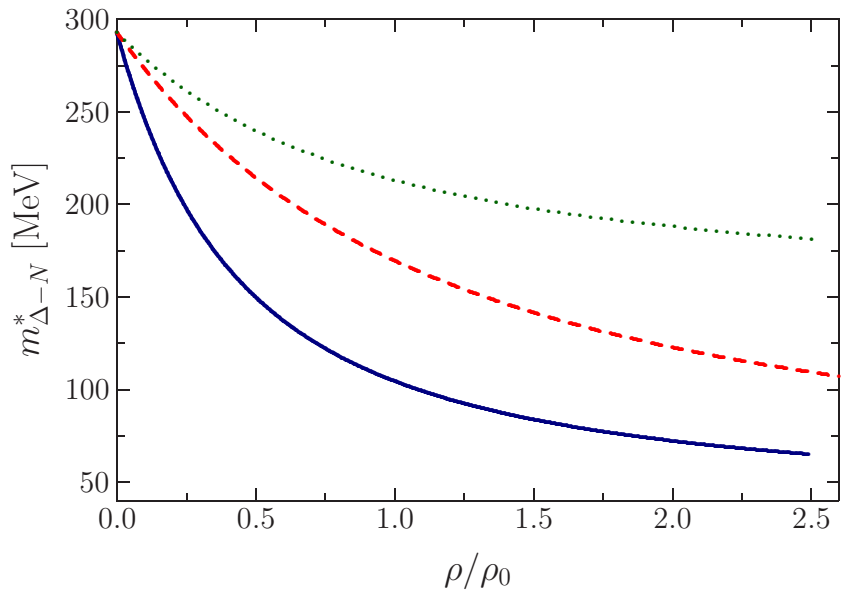}}
                \caption{(Color online). The density dependence of the
                  $\Delta- N$ mass difference in nuclear matter. The
                  notations and input parameters are the same as
                  those in Fig.~\ref{fig:1}.}
                \label{fig:2}
\end{figure}

In Fig.~\ref{fig:2}, the dependence of the $\Delta- N$ mass difference
on the nuclear matter density is drawn.
The results show that the modified Skyrme term leads to rather
different results from that without the modification. With the
Skyrme term modified (see solid and dashed curves), the results of
$m_{\Delta- N}^*$ fall off faster as the density increases in
comparison with that with the original Skyrme term (see dotted
curve). Of course, this is due to the explicit density dependence of
the moment of inertia~(\ref{inmom}) through the additional density
functional $\gamma$. It implies that it is easier to make the nucleon
excited to the $\Delta$ state in nuclear matter, which seems more
realistic than that without the modification of the Skyrme term.

\begin{figure}[htb]
\centerline{\includegraphics[scale=0.8]{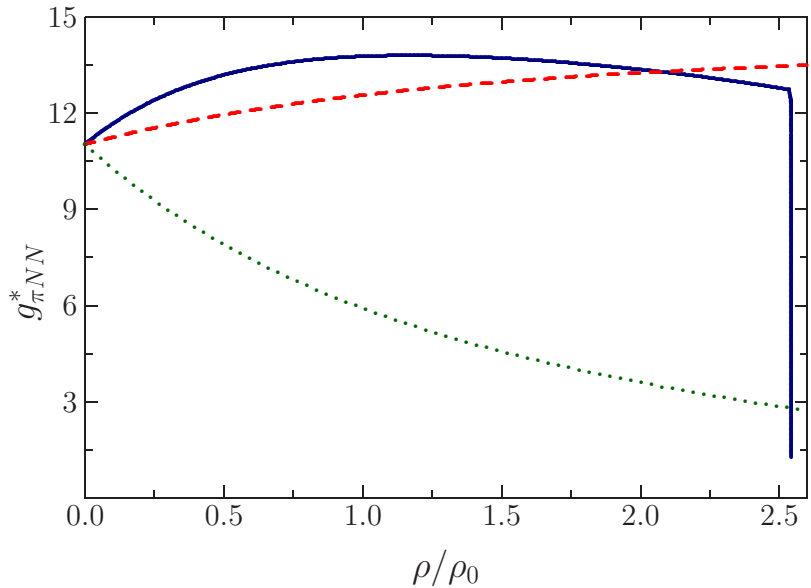}}
                \caption{(Color online). The dependence of $\pi NN$
                  coupling constant on the density. Notations are
                  similar to those in Fig.~\ref{fig:1}. }
                \label{fig:3}
\end{figure}

In Fig.~\ref{fig:3}, the changes of the $\pi NN$ coupling constant are
depicted.
Here the modifications of the Skyrme term bring about more dramatic
results. When the Skyrme term is intact, i.e. $\gamma=1$, $g_{\pi
  NN}^*$ monotonically decreases as the density increases (see the
dotted curve in Fig.~\ref{fig:3}). However, when one introduces the
density dependence of the Skyrme term, the results are noticeably
changed. Using the parametrization of Eq.~(\ref{gamma}) with the
different values of input parameter $c_0$, we find that
the $\pi NN$ coupling in medium changes drastically. For
example, with the value of $c_0=0.09m_\pi^{-3}$ the in-medium
pion-nucleon coupling constant $g_{\pi NN}^*$ starts to increase
monotonically up to high ($\rho\sim 5\rho_0$) densities as drawn in
the dashed curve. The $g_{\pi NN}^*$ with $c_0=0.09m_\pi^{-3}$ will
disappear at around $\rho\sim 5\rho_0$.
On the other hand, if one uses $c_0=0.21m_\pi^{-3}$, it is getting
increased up to the normal nuclear matter density and stays more or
less constant. Then it slowly falls off as the density increases. When
it approaches the critical point $\rho\sim 2.54\rho_0$, it drops
sharply and goes to zero (see the solid curve in
Fig.~\ref{fig:3}). Above the critical point ($\rho > \rho_{\rm
  crit}\approx 2.54\rho_0$), the skyrmion does not exist. It is not
surprising, because the sign of the combination $\alpha_p\gamma$ in
Eq.~(\ref{feq}) is changed and therefore there is no stable solitonic
solution anymore.

Let us draw the attention again to the bulk properties of nuclear
matter in order to understand the modifications of the Skyrme term
better. Following Klebanov~\cite{Klebanov:1985qi}, after 
quantizing and using the formula presented in Ref.~\cite{Lee:2010sw},
one can estimate the symmetry energy in the semiempirical formula for 
the nuclear binding energy:
\begin{equation}
  \label{eq:11}
E_{\rm sym}=\frac{1}{12}\,m_{\Delta- N}^*\,.
\end{equation}
This crude formula of the symmetry energy already provides
the enlightening results. For example, $\gamma$ parameterized as in
Eq.~(\ref{gamma}), $E_{\rm sym}(\rho_0)\approx 14.19$~MeV for
$c_0=0.09m_\pi^{-3}$ whereas $E_{\rm  sym}(\rho_0)\approx 8.71$~MeV
for $c_0=0.21m_\pi^{-3}$. These results for the symmetry energy
must be compared with the experimental one $E_{\rm sym}\sim
20 - 30$~MeV. The order of the symmetry energy calculated
within the Skyrme model is comparable to the experimental data. To
estimate the symmetry energy more accurately, however, one should
consider the minimization of the whole binding energy taking into
account the interplay between the different terms in the mass formula
as we stated already. Moreover, one should consider the effects of
finite nuclei and explicit isospin-breaking effects and so on. We want
to mention that this is also possible within the in-medium modified
Skyrme model and can be done as in
Refs.~\cite{Meissner:2007ya,Meissner:2008mr},
an additional modification of the Skyrme term being performed as
was done in the present work.

\section{Summary and Outlook \label{Sum}}
In the present investigation, we aimed at studying the modifications
of the quartic term in the Skyrme model. The results from this work
shows that the additional modifications change dramatically the
whole picture and allows one to understand the role of the
modifications in a more comprehensive way. One can note that an
alternative approach to many-body systems within the Skyrme
model~\cite{Battye:2006na,Manton:2006tq} points to the changes of
the input parameters (so called ``calibration'') according to the number
of baryons in the system. Within our approach these changes were shown in
a more realistic and transparent way and were treated not only at the level
of the system but also at the level of its constituents.

From the previous studies we know that the large renormalization of
the nucleon mass in nuclear medium causes one of the difficulties to
bind the infinite nuclear matter~\cite{Yakhshiev:2004pj} and to
reproduce the correct values of the Nolen-Schiffer anomaly in mirror
nuclei within the in-medium modified Skyrme
model~\cite{Meissner:2007ya}. The relatively small change of the
nucleon mass in nuclear matter within the present approach is an
interesting result, which can be used to reproduce the correct value
of the Nolen-Schiffer anomaly. In addition to the small nucleon mass
renormalization, the dramatic changes in the pion-nucleon coupling
constant gives the opportunity to revise the previous investigation on
nuclear matter related to the quantum-mechanical many-body
problems~\cite{Yakhshiev:2004pj}.

More completely, one can improve the present approach by
treating the density of the system in a fully consistent
way. Since it is a sum of the single skyrmion densities in a given
initial configuration and one can study the equilibration of the
matter to its ground state, taking into account the deformation
effects on the properties of its constituents~\cite{Meissner:2008mr}.

Having considered the modification of the Skyrme term, we assert
that the present approach is self-consistent: it treats the single
hadron properties, the hadron-hadron interactions and the bulk matter
properties on an equal footing. Moreover, it is closely related to
phenomenological low-energy data at the single-hadron level as well as
at the level of hadronic systems. The present approach can be extended to
the studies of the properties of finite nuclei and its constituents.

\section*{Acknowledgments}
The authors are grateful to H.K. Lee for the discussion of the
symmetry energy within the Skyrme model. The present work is
supported by Basic Science Research Program through the National
Research Foundation of Korea (NRF) funded by the Ministry of
Education, Science and Technology (grant number: 2009-0089525).

\end{document}